\documentclass[final]{IEEEtran}
\usepackage{amsthm,amssymb,graphicx,multirow,amsmath,color,amsfonts}
\usepackage[update,prepend]{epstopdf}
\usepackage[noadjust]{cite}
\usepackage[latin1]{inputenc}
\usepackage{tikz}
\usepackage{bbm} 
\usepackage{pdfpages}
\usepackage{tabulary}
\usepackage{multirow}
\usepackage{comment}
\usepackage{textcomp}
\usepackage{xcolor}
\usepackage{adjustbox}
\usepackage[center]{caption}

\usepackage{dsfont}

\def\nb0{{\mathbf{0}}}
\def\nb1{{\mathbf{1}}}

\newtheorem{lemma}{Lemma}

\newtheorem{definition}{Definition}

\newtheorem{remark}{Remark}

\allowdisplaybreaks 

\usepackage{setspace}	

\begin{document}
\pagenumbering{gobble}
\graphicspath{{./Figures/}}
\title{
Stochastic Geometry-based analysis of Airborne Base Stations with Laser-powered UAVs}
\author{
Mohamed-Amine Lahmeri, Mustafa A. Kishk, and Mohamed-Slim Alouini
\thanks{Authors are with Computer, Electrical, and Mathematical Science and Engineering (CEMSE) Division,
King Abdullah University of Science and Technology (KAUST), Thuwal, Makkah Province, Saudi Arabia (e-mail:\{mohamed.lahmeri ; mustafa.kishk; slim.alouini\}@kaust.edu.sa).} 

}

\maketitle 
\begin{abstract}
One of the most promising solutions to the problem of limited flight time of unmanned aerial vehicles (UAVs), is providing the UAVs with power through laser beams emitted from Laser Beam Directors (LBDs) deployed on the ground. In this letter, we study the performance of a laser-powered UAV-enabled communication system using tools from stochastic geometry. We first derive the energy coverage probability, which is defined as the probability of the UAV receiving enough energy to ensure successful operation (hovering and communication). Our results show that to ensure energy coverage, the distance between the UAV and its dedicated LBD must be below a certain threshold, for which we derive an expression as a function of the system parameters. Considering simultaneous information and power transmission through the laser beam using power splitting technique, we also derive the joint energy and the Signal-to-noise Ratio (SNR) coverage probability. The analytical and simulation results reveal some interesting insights. For instance, our results show that we need at least 6 LBDs/10km$^2$ to ensure a reliable performance in terms of energy coverage probability.

\end{abstract}
\section{Introduction} \label{sec:intro}
Due to its wide spectrum of applications and use cases, UAVs have recently known an unprecedented spread and attracted research interests worldwide~\cite{added3}. UAVs' main attractions include flexible deployment, high maneuverability, and the continuous decrease in their cost. These advantages motivated integrating UAV-mounted BSs into existing cellular networks to improve coverage and capacity~\cite{integrating}, while taking advantage of the maneuverability of the UAVs to optimize its path~\cite{legpp,Moving}. However, the feasibility and the reliability of the UAV-enabled applications still face some crucial challenges, especially for long-duration missions, due to the limited energy resources on-board~\cite{added1,added3}. Unfortunately, the majority of the commercially available UAVs struggle to stay in the air for more than half an hour. Thus, the UAV is always obliged to abort its mission to refuel or change its battery. 
To overcome this technical challenge, many works in the literature focused on optimizing the consumption of the on-board energy by increasing the battery capacity~\cite{battery}, optimizing the UAV trajectory~\cite{7888557}, or optimizing the UAV placement~\cite{7918510}. While these solutions maximize the efficiency of the energy consumption on-board, they can not provide the significant increase in the UAV flight time, which is required for many UAV-enabled applications.

In this context, a novel technology to prolong mission duration is laser beaming~~\cite{SLIPT,6,8648453}. This technology proved the ability to enable much longer UAV flight times and is currently being developed by a number of companies~\cite{lasermotive}. Implementing a high power laser device is proven to be possible using an appropriate power beaming system, in which an energy-rich laser array can be oriented through a complex optical system (set of mirrors or diamonds) and then shines on the target UAV. More interestingly, the laser beam could be used for both energy harvesting and information transfer.

In this letter, we use tools from stochastic geometry to model and analyze the performance of laser-powered UAV-mounted BSs. In particular, we consider simultaneous information and power transmission from the LBD to the UAV for (i) charging the UAV and (ii) providing the UAV with wireless backhaul link through the laser beam. More details about the contributions in this paper are provided next.

\textit{Contributions}. Modeling the locations of the LBDs as a homogeneous Poisson Point Process (PPP), we focus on a typical UAV that is harvesting power and receiving information from its nearest LBD simultaneously using power splitting technique. We first derive a closed-form expression for energy coverage probability using the Free-space Optical (FSO) range equation. We also study the effect of atmospheric turbulence on this probability. Next, we evaluate the SNR coverage probability at the UAV, which is the probability of successful communication through the backhaul link. Furthermore, we derive the joint probability of energy and SNR coverage and exploit it to draw useful system-level insights on the influence of some system parameters on the performance such as the power splitting factor and the LBD density. We provide general expressions for all the derived performance metrics, which can be applied using the energy consumption models for either fixed-wing or rotary-wing UAV.
\section{System Model} \label{sec:SysMod}
\begin{figure}
    \centering
    \includegraphics[width=0.3\textwidth]{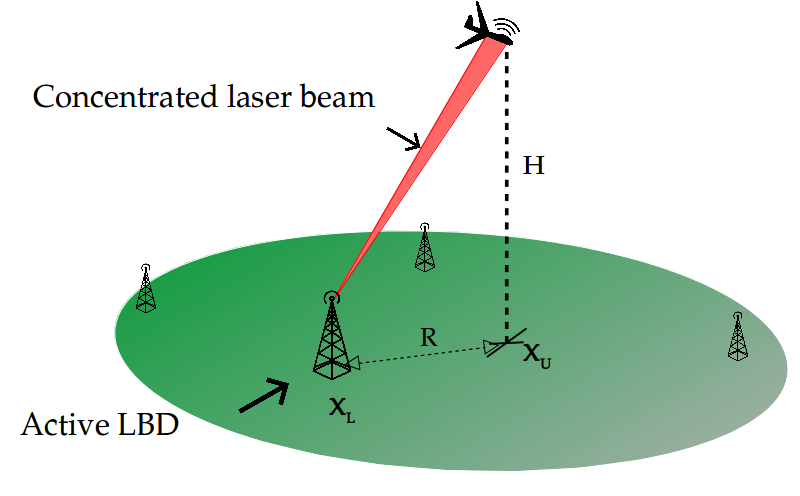}
    \caption{System description of a laser-powered UAV.}
    \label{fig:2}
\end{figure}
We consider a UAV-enabled communication system, where a UAV is providing cellular coverage for ground users. The UAV is powered by LBDs that are deployed on the ground
and spatially distributed according to a Poisson Point Process (PPP), $\Phi_L=\{x_i\}\in \mathbb{R}^2$ with density $\lambda_L $.
We assume that the high altitude of the UAVs enables them to effectively establish Line-of-sight (LOS) link with the LBDs. All LBDs transmit energy with the same fixed power $p_{trans}>0$ and the UAV associates with its {\em nearest LBD}.

As shown in Fig.~\ref{fig:2}, we suppose that the UAV flies with a constant altitude $H>0$. Without loss of generality, we focus on a typical UAV located at $(0,0,H)$. The serving LBD is located at $(\mathbf{x_{L}},0)$; $\mathbf{x_{L}} \in \mathbb{R}^2$. Consequently, the distance between the UAV and its serving LBD is $ d =\sqrt{R^2+H^2} $, where $R=\vert\vert x_L \vert\vert$ denotes the distance between the UAV's projection in $\mathbb{R}^2$ and its serving LBD. Given that the locations of the LBDs are modeled by a PPP, the location of the serving LBD, and consequently R, are random variables.

In addition to energy harvesting and communication with users on the ground, the UAV is also backhauling with its serving LBD through intensity modulation. In this work, we focus on downlink. Hence, the UAV is considered as a receiver when communicating with its serving LBD. The UAV adopts power splitting technique which divides the laser power received from the LBD into two streams: (i) energy harvesting with power ratio $1-\delta_s$ and (ii) backhaul link information with power ratio $\delta_s$.

\subsection{Power Harvesting Model}
The energy harvesting through the laser link can be derived using the commonly known FSO range equation. We suppose a linear energy harvesting model with an efficiency $w$ and, thus, the power harvested at the UAV, when associating with an LBD at distance $R$ from its projection in $\mathbb{R}^2$, is represented by~\cite{killinger}: 
\begin{align} \label{eq1}
  p_{\rm harv}(R) & = (1-\delta_s)p_{\rm rec}(R)  = \frac {(1-\delta_s) \omega A\chi p_{trans} e^{-\alpha d}}{(D+d\Delta\theta)^{2}}\nonumber\\&=\frac {(1-\delta_s) \omega A\chi  p_{trans} e^{-\alpha \sqrt{R^2+H^2}}}{(D+\sqrt{R^2+H^2}\Delta\theta)^{2}},
\end{align} 
where $p_{\rm rec}(R)$ is the total power received from the serving LBD, $(1-\delta_s)$ is the fraction of the received power dedicated to energy harvesting, $A$ is the area of the receiver telescope or collection lens, $D$ is the size of the initial laser beam, $\alpha$ is the attenuation coefficient of the medium, $\chi$ is the combined transmission receiver optical efficiency, $\Delta\theta$ is the angular spread of the laser beam. The angular spread $\Delta\theta$ is equal to $\frac{D_d}{f}$, where $D_d$ is the size of the detector and $f$ its focal length.
 
Note that (\ref{eq1}) is derived from the Beer-Lambert equation. It takes into account the scattering effect and the divergence of the laser beam when propagating into the atmosphere but does not include the effect of the atmospheric turbulence.
\subsection{Turbulence Effect}
The received power can be affected by turbulence. In the literature, many statistical models were proposed to model the intensity fluctuation caused by turbulence. A Log-Normal distribution can catch the turbulence effect in the weak-to-moderate regime. However, for moderate-to-strong turbulence regime, a Gamma-Gamma distribution can be used. Although we assume a Log-Normal distribution in the simulation results section, we keep our analytical results general for any kind of distribution. The probability density function of the Log-Normal distribution is given by:
\begin{align}
f_{h_t}(h_t) = \frac{1}{2h_t\sqrt{2 \pi \sigma^2}}exp\Big(-\frac{({\rm ln}(h_t)+2\sigma^2)^2}{8\sigma^2} \Big),
\end{align}
where $h_t$ is the random variable that represents the turbulence effect, and $\sigma^2$ is the variance, which is given by: 
\begin{align*}
\sigma^2=0.3k^{\frac{7}{6}}C_n^2(H) R^{\frac{11}{6}},
\end{align*}
where $C_n^2(H)$ is the index of refraction structure parameter at altitude $H$ and $k$ is the optical wavenumber.

\subsection{Power Consumption Model}
The UAV's power consumption is composed of the propulsion power $p_{\rm prop}$ and the communication-related power $p_{\rm comm}$. The value $p_{\rm comm}$ represents the power consumed by the payload of the UAV for communication and on-board processing. The propulsion power differs with the type of the used UAV. We provide expressions for fixed-wing and rotary-wing UAV propulsion power below.
\subsubsection{Fixed-wing UAV}
The instantaneous power consumed by the fixed-wing UAV is represented by the following equation~\cite{6}:
\begin{align}
  p_{\rm prop} & =\Bigg|c_1\left\|\ v \right\|^3+\frac {c_2}{\left\|\ v \right\|}(1+\frac {\left\|\ a \right\|^2-\frac{(a^{T}v)^2}{v^2} }{g^2})+ma^T v\Bigg|,
\end{align} 
where $v$ denotes the instantaneous UAV velocity, $a$ denotes the UAV acceleration, $m$ is the mass of the UAV, $c_1$ and $c_2$ are two parameters related to the aircraft's weight, wing area, and air density.

\subsubsection{Rotary-wing UAV}
The propulsion power for a rotary-wing UAV can be approximated using~\cite{rotary}: 
\begin{align}
p_{\rm prop}= P_0 \bigg( 1+\frac{3v^2}{U_{\rm tip}^2}\bigg)+ \frac{P_i v_0}{v}+\frac{1}{2}d_0 \rho s\mathcal{A}v^3,
\end{align}
where $U_{\rm tip}$ denotes the top speed of the rotor blade,  $v_0$ is the mean rotor induced velocity, $d_0$ is the fuselage drag ratio, $s$ is  rotor solidity, $\rho$ denotes the air density and $\mathcal{A}$ denotes the rotor disc area. $P_0$ and $P_i$ are two constants related to UAV weight, rotor radius, blade velocity, and other system parameters.

In this work, our first objective is to derive the energy coverage probability, which is formally defined below.

\begin{definition}[Energy Coverage Probability] \label{def:1}
The energy coverage probability at the UAV is defined as the probability that the harvested power is greater than the consumed power:
\begin{align}\label{eqn:def_energy}
P_{\rm energy}=\mathbb{P}(p_{\rm harv}(R)> p_{\rm prop}+p_{\rm comm}).
\end{align} 
\end{definition}   
\subsection{Backhaul Link}
To communicate with the UAV, the LBD uses laser intensity modulation. We assume that the information transfer is done using On-Off keying modulation (OOK).
The UAV is equipped with a direct detection (DD) system that responds only to the instantaneous power of the collected field. The DD system is composed of a receiving lens that focuses the optical beam onto a photo-detecting surface, which converts the instantaneous power into an electrical signal for processing. Taking into account the atmospheric effects of turbulence, the responsivity of the detector, and the splitting factor, the average photocurrent collected is given by: 
\begin{align}
i_s= \delta_s \eta h_t p_{\rm rec}(R),
\end{align}
where $\eta$ is the photodiode responsivity, $\delta_s$ is the power splitting factor, and $h_t$ represents the turbulence effect on the received signal. Since we are using a high laser power, which wavelength is in the infrared domain, we adopted a shot noise limited regime where the SNR expression for the FSO link can be expressed as follows~\cite{SNR,killinger}:

\begin{align}\label{snr_equ}
{\rm SNR}_{\rm UAV} =\frac{ i_s^2}{\sigma_n^2}=\frac{\delta_s^2 \eta^2 h_t^2 p_{\rm rec}(R)^2}{2h\nu i_s\Delta f }=\frac{\delta_s \eta h_t p_{\rm rec}(R)}{2 h \nu \Delta f},
\end{align}
where $h$ is Planck\textquotesingle s constant, $\nu=\frac{c}{\lambda}$ is the photon\textquotesingle s frequency, $\lambda$ is the wavelength, $c$ is the speed of light, and $\Delta f $ is the modulation frequency bandwidth. 
Using the SNR expression in (\ref{snr_equ}), we define  the SNR coverage probability as follows.

\begin{definition}[SNR Coverage Probability]Given that the power splitting technique is adopted, SNR coverage probability is defined as the probability that the information-carrying part of the received signal from the LBD at the UAV can achieve a target SNR threshold $\beta$:  \label{def:2}
\begin{align}\label{snr_def}
P_{\rm SNR}=\mathbb{P}({\rm SNR_{\rm UAV}}> \beta).
\end{align}
\end{definition}
The ultimate objective from the analysis provided in this paper is to find the joint SNR and energy coverage probability, which is defined next.
\begin{definition}[Joint Coverage Probability]The joint coverage probability is the probability that the UAV is able to successfully receive and decode the information, while harvesting enough power for propulsion and communication:  \label{def:1}
\begin{align}\label{eqn:joint_def}
P_{\rm joint}=\mathbb{P}(p_{\rm harv}(R)> p_{\rm prop}+p_{\rm comm} , {\rm SNR}_{\rm UAV}> \beta).
\end{align}
\end{definition}
\section{Performance Analysis}
In this section, our aim is to derive the previously defined performance metrics: (i) the energy coverage probability, (ii) the SNR coverage probability, and (iii) the joint coverage probability. 
We first derive the energy coverage probability in the case of no turbulence, in order to provide some insightful expressions. We will incorporate the turbulence effect in the energy coverage probability later in this section.
\begin{lemma}[Energy Coverage Probability]\label{lem:1}
The energy coverage probability in the case of no turbulence is given by: 
\begin{align}
\label{4}
P_{\rm energy}=1-e^{-\lambda_L\pi R^{*2}},
\end{align}
where $R^*$ is given by  
\begin{small}
\begin{align}\label{eqn:r_star}
 R^*=\Bigg[ \Bigg\{\frac{2}{\alpha}W_0\Bigg(\frac{\alpha}{2\Delta\theta }\sqrt{\frac{(1-\delta_s)\omega A\chi  p_{trans} e^{\frac{\alpha D}{\Delta\theta }}}{p_{\rm prop}+p_{\rm comm}}}\Bigg)-\frac{D}{\Delta\theta}\Bigg\}^2-H^2\Bigg]^{\frac{1}{2}},
\end{align}
\end{small}where $W_0(.)$ is the principal branch of the Lambert W function~\cite{corless1996lambertw}.
\end{lemma}
\begin{IEEEproof}
See Appendix~A.
\end{IEEEproof}
\begin{remark}
Agreeing with intuition, the probability of having enough harvested power for the UAV is an increasing function of the density of the LBDs existing in the ground. Furthermore, the value of $R^{*}$ represents a threshold on the distance between the UAV and the LBD that ensures energy coverage. The provided expression in (\ref{eqn:r_star}) for $R^{*}$ captures the effect of all the system parameters on the energy coverage probability, such as the UAV altitude $H$, the propulsion power $p_{\rm prop}$, and the communication-related power $P_{\rm comm}$,
\end{remark}
Next, we derive the energy coverage probability under atmospheric turbulence effect with general probability distribution.
\begin{lemma}[Effect of Turbulence on the Harvested Power]\label{lem:2}
Taking into account the effect of atmospheric turbulence, the energy coverage expression is given by:
\begin{align}
P_{\rm energy}= \int_{0}^{\infty}  (1-F_{h_t}(a(r)))2\pi\lambda_L r e^{-\lambda_L \pi r^2} {\rm d}r,
\end{align}
where $F_{h_t}$ is the cumulative distribution function of $h_t$, and $a(r) = \frac{p_{\rm prop}+p_{\rm comm}}{p_{\rm harv}(r)}$.
\end{lemma}

\begin{IEEEproof}
Incorporating the turbulence effect into the energy coverage probability definition in (\ref{eqn:def_energy}), we have  
\begin{align}
&P_{\rm energy}=\mathbb{P}( h_t p_{\rm harv}(R)  > p_{\rm prop}+p_{\rm comm})\nonumber\\&= \mathbb{E}_R \left[\mathbb{P}\left( h_t p_{\rm harv}(R)  > p_{\rm prop}+p_{\rm comm}\bigg| R=r\right ) \right ] \nonumber\\ &=  \int_{0}^{\infty}  \mathbb{P}\left( h_t p_{\rm harv}(R)  > p_{\rm prop}+p_{\rm comm}\bigg| R=r\right ) f_R(r) {\rm d}r\nonumber \\ &= \int_{0}^{\infty}  (1-F_{h_t}(a(r)))2\pi\lambda_L r e^{-\lambda_L \pi r^2} {\rm d}r,
\end{align}
where the probability distribution of the distance between the UAV's projection in $\mathbb{R}^2$ and its nearest LBD is given by $f_R( r)= 2 \pi r \lambda_L e^{-\lambda_L \pi r^2}$. 
\end{IEEEproof} 
In addition to the power transfer, the LBD transmits information to the UAV through the laser beam via intensity modulation. We derive the SNR coverage probability at the UAV for a general distribution of $h_t$ in the following lemma.
\begin{lemma}(SNR Coverage Probability)\label{snr_lem}
The SNR coverage probability is given by:
\begin{align}
{P}_{\rm SNR}=\int_{0}^{\infty}  \Bigg(1-F_{h_t}(b(r))\Bigg)2\pi\lambda_L r e^{-\lambda_L \pi r^2} {\rm d}r,
\end{align}
where $b(r) = \frac{2 q \Delta f \beta} {\delta_s\eta p_{\rm harv}(r)}$.
\end{lemma}

\begin{IEEEproof}
Using the same approach in the proof of Lemma~\ref{lem:2}, and substituting in (\ref{snr_equ}) and (\ref{snr_def}), the final result in Lemma~\ref{snr_lem} follows. 
\end{IEEEproof}
Using the results in Lemma~\ref{lem:2} and Lemma~\ref{snr_lem}, the joint coverage probability, defined in Definition~\ref{def:1}, is derived next.
\begin{lemma}(Joint Coverage Probability) The joint coverage probability is given by: 
\label{lem:4}
\begin{align}\label{eqn:joint_lem}
{P}_{\rm joint}&= {P}_{\rm energy}\mathds{1}\left( \mathcal{K}>1\right)+{P}_{\rm SNR}\mathds{1}\left( \mathcal{K}\leq 1\right),
\end{align}
where $\mathcal{K}=\frac{\eta\delta_s(p_{\rm prop}+p_{\rm comm)}}{2h\nu \Delta f \beta(1-\delta_s)}$ and $\mathds{1}(.)$ is the indicator function.
\begin{IEEEproof}
Using the definition in (\ref{eqn:joint_def}), we have
\begin{small}
\begin{align}
{P}_{\rm joint}&= \mathbb{P}((1-\delta_s)h_t p_{\rm rec}(R)\geq p_{\rm prop}+p_{\rm comm} ,  \frac{\delta_s\eta h_t p_{\rm rec}(R)}{2 h \nu \Delta f}  \geq \beta)\nonumber\\ &=\mathbb{P}(h_t p_{\rm rec}(R)\geq \frac{p_{\rm prop}+p_{\rm comm}}{1-\delta_s} ,  h_t p_{\rm rec}(R)\geq \frac{2 h \nu \Delta f \beta}{\eta\delta_s}),\nonumber
\end{align}
\end{small}
\noindent which leads to the final result in (\ref{eqn:joint_lem}).
\end{IEEEproof}
\end{lemma}
\begin{remark}\label{remark2}
From (\ref{eqn:joint_lem}), we can easily note that the joint coverage probability is either dominated by the energy coverage $P_{\rm energy}$ or by the SNR coverage $P_{\rm SNR}$, depending on the value $\mathcal{K}=\frac{\eta\delta_s(p_{\rm prop}+p_{\rm comm)}}{2h\nu \Delta f \beta(1-\delta_s)}$. The term $\mathcal{K}$ captures the effect of multiple system parameters such as the $\beta$ and $\delta_s$. 
\end{remark}
\section{Numerical Results and Simulations}
\begin{figure}
    \centering
    \includegraphics[width=0.87\columnwidth]{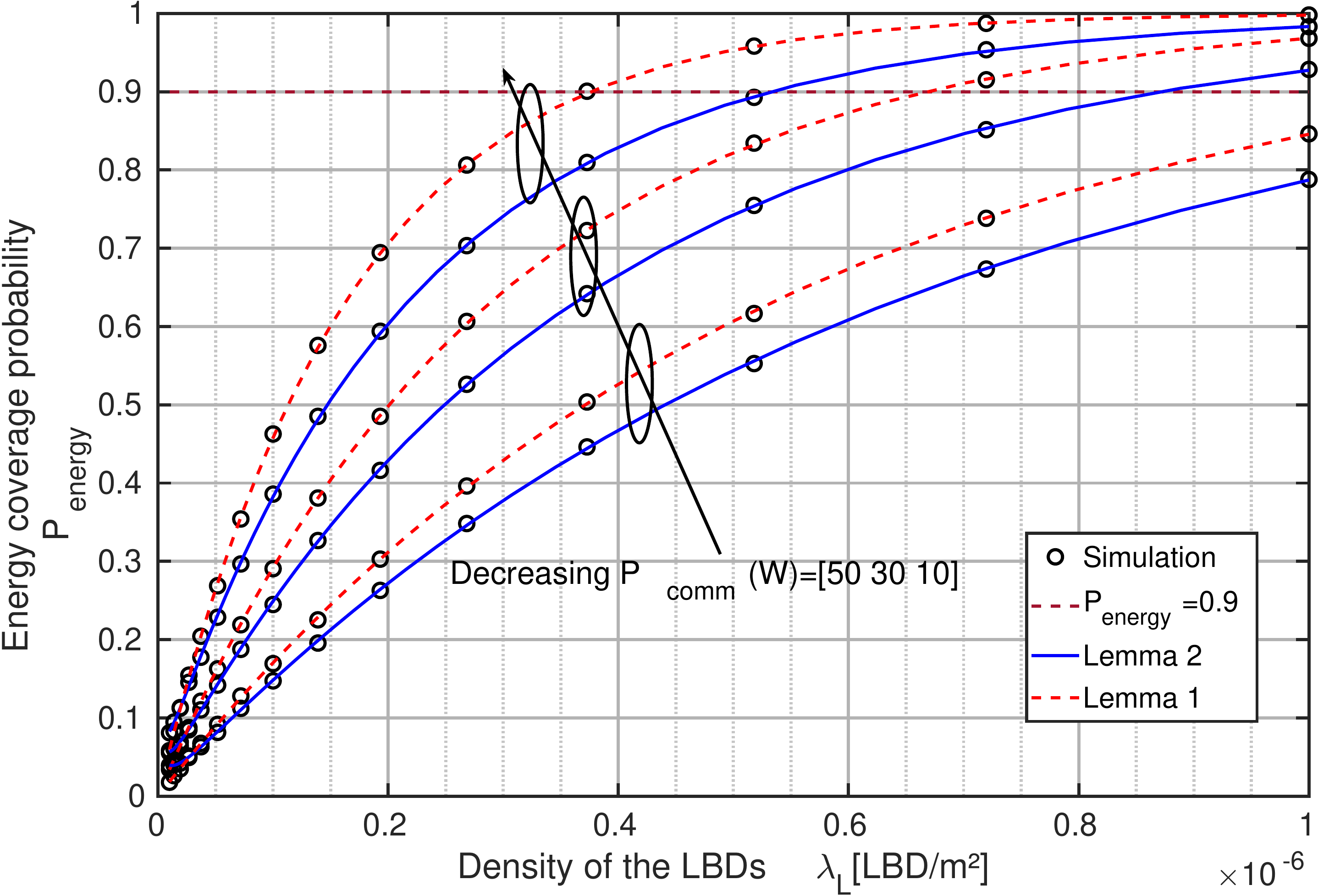}
    \caption{Energy coverage for a Laser-powered UAV.}
    \label{fig:PENE}
\end{figure}

In this section, we verify the theoretical results derived throughout the paper using Monte Carlo simulations. In addition, we draw some useful system-level insights and provide some comments and recommendations for an efficient design of a laser-powered UAV-enabled communication system. The values of the system parameters considered in the simulation setup comply with the values provided by commercially available laser-powered UAVs, and are summarized in Table~\ref{table:1}. In Fig.~\ref{fig:PENE}, we plot the energy coverage probability $P_{\rm energy}$ against different values of $\lambda_L$. The simulation results confirm the accuracy of the theoretical results we proposed in Lemma~\ref{lem:1} and Lemma ~\ref{lem:2}, which was derived without any approximations. The red dashed lines correspond to the case of no turbulence effect, while the continuous blue lines correspond to the energy coverage under a Log-Normal distributed atmospheric turbulence. 

In order to extract a useful threshold on the density of the LBDs, we impose to satisfy a stringent minimum on  $P_{\rm energy}$ which is 90\% probability.  For instance, this level corresponds to $\lambda_L= 0.52\times10^{-6}$ LBD/m$^2$, in the case of $P_{\rm comm}=10 W$ and with the presence of weak turbulence. Putting it into words, we recommend having 6 LBDs each 10 km$^2$ to ensure energy coverage. This level of ${P}_{\rm energy}$ is safe, because even in the worst cases where the UAV is cut off from its serving LBD, a backup battery could solve this problem efficiently until the UAV is able to establish a charging link with a new LBD.
\begin{table}[]\caption{System parameters~\cite{6,lasermotive,Outage}}
\centering
\begin{adjustbox}{width=\columnwidth,center}
\begin{tabular}{|c|c|c|c|}

\hline
Parameter & Value & Parameter & Value \\ \hline
Area         & [300 {km}, 300 {km}]$^2 $   & $\#$ of    iterations         & 10000     \\ \hline
H         & 100 m    & UAV type         & Fixed-wing     \\ \hline
D         &0.1 m       & $c_1$          & $9.26\times10^4$ kg/m       \\ \hline
$\Delta \theta$ &$3.4\times10 ^{-5}$        &$c_2$           &$2250$ kgm$^3$/s$^4$      \\ \hline
$\alpha$          & $10^{-6}$ m       &$wA\chi$ &$0.004$ m$^2$       \\ \hline
$p_{trans}$          &$600$ W      &$C_n^2$           &$0.5\times10^{14}$       \\ \hline
$\eta  $        & $0.5$ A/W     &$h$          &$6.63\times10^{-34} $m$^2$kg$s^{-1}$  \\ \hline
$ C$  &$0.004$ m$^2$ &$\Delta f $                &1 GHz                \\ \hline
$\delta_s$                 &$10^{-5} $          &$k$   &$5.92\times 10^6$ nm$^{-1} $   \\ \hline
$\lambda$                 &$0.785$ $\mu$m           &$v$   &$30$ m/s   \\ \hline
$P_{prop}$                 &100 W           &$a$   &$0$ m/s$^2$   \\ \hline
\end{tabular}
\end{adjustbox}
\label{table:1}
\end{table}

\begin{figure}
    \centering
    \includegraphics[width=0.9\columnwidth]{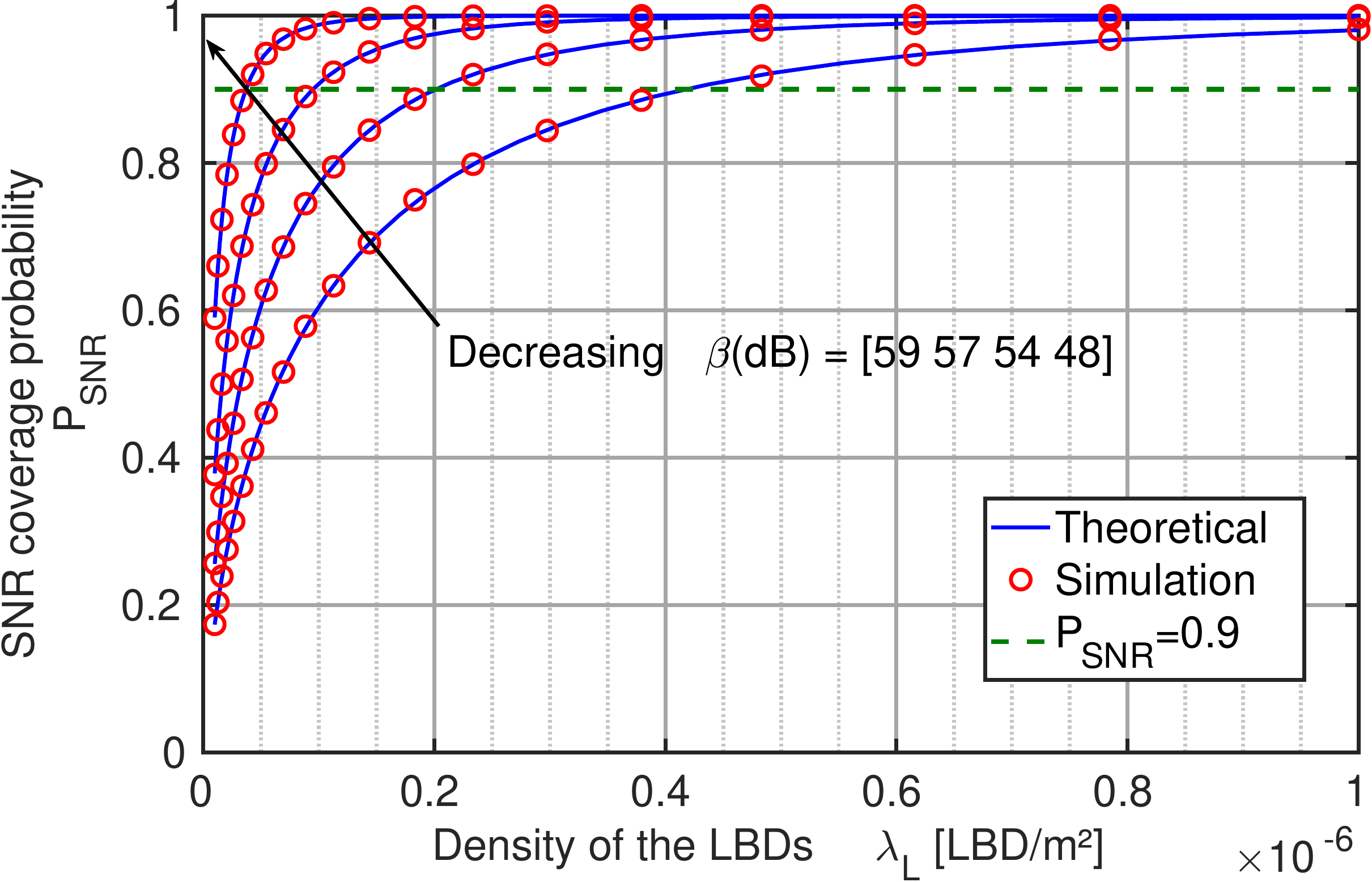}
    \caption{SNR coverage probability under Log-Normal distributed turbulence.}
    \label{fig:SNR}
\end{figure}

Fig.~\ref{fig:SNR} shows the SNR coverage for different values of $\beta$. 
Comparing Fig.~\ref{fig:PENE} and~\ref{fig:SNR}, we note that much less LBDs are needed to ensure SNR coverage than the number of LBDs needed to ensure energy coverage. In other words, FSO communication can provide coverage for ranges longer than Laser power transmission.

\begin{figure}
    \centering
    \includegraphics[width=0.9\columnwidth]{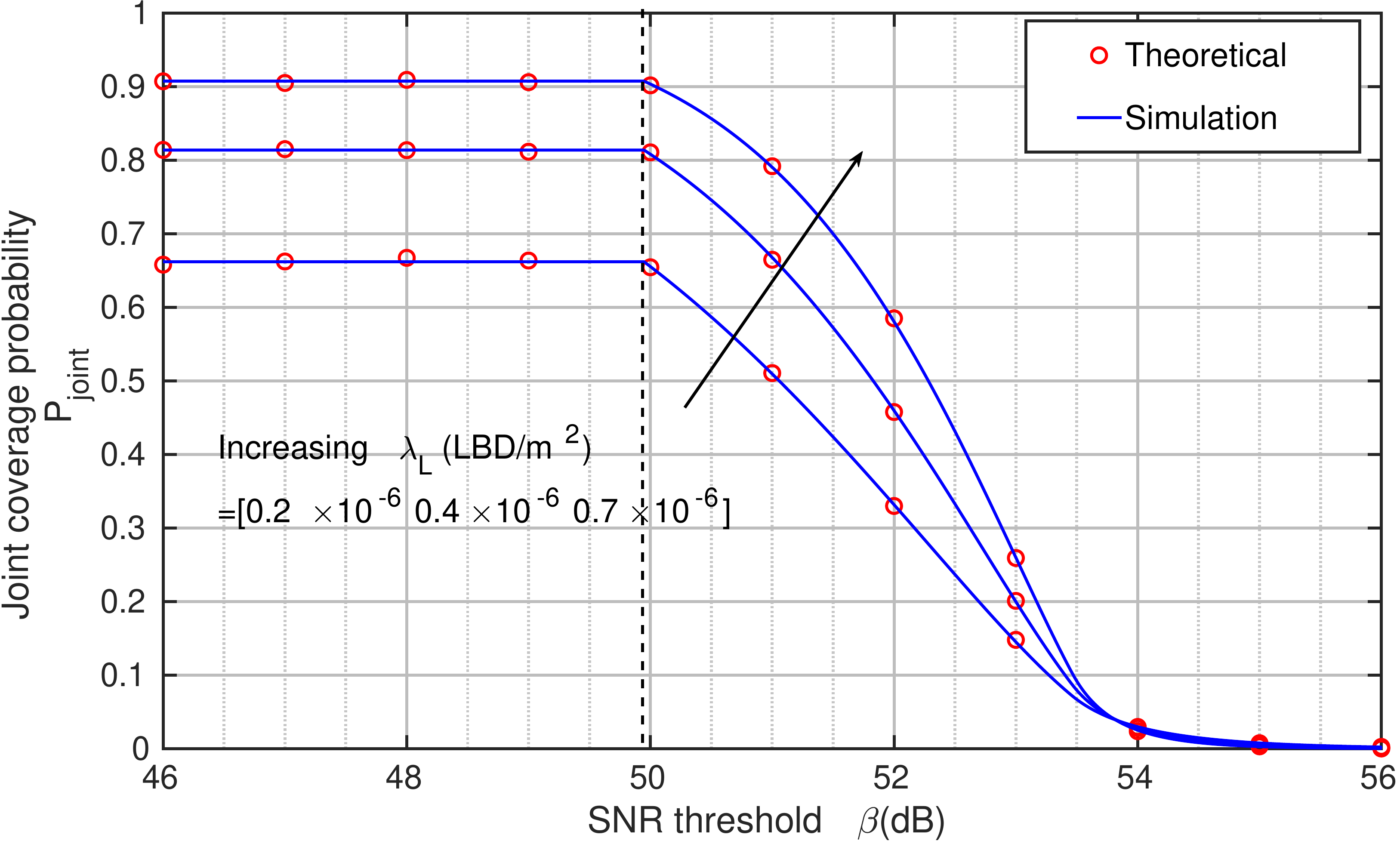}
    \caption{Joint coverage probability against SNR Threshold $\beta$, $\delta_s=10^{-6}$.}
    \label{fig:PJOINT1}
\end{figure}

In Fig~\ref{fig:PJOINT1}, we observe that $\mathcal{K}$, defined in Lemma~\ref{lem:4}, is greater than one as long as $\beta$ is less than 50dB. This makes the joint coverage probability dominated by the energy coverage and, thus, it remains constant at this range of values of $\beta$.  When $\beta$ is greater than 50dB, the joint coverage probability starts to decrease as it is dominated by the SNR coverage. 
 These results verify our comments in Remark~\ref{remark2} that the joint coverage probability $P_{\rm joint}$, provided in Lemma~\ref{lem:4}, is dominated either by the energy coverage probability $P_{\rm energy}$ or by the SNR coverage probability $P_{\rm SNR}$, depending on the value of $\mathcal{K}$, which is a function of $\beta$.

\begin{figure}
    \centering
    \includegraphics[width=0.9\columnwidth]{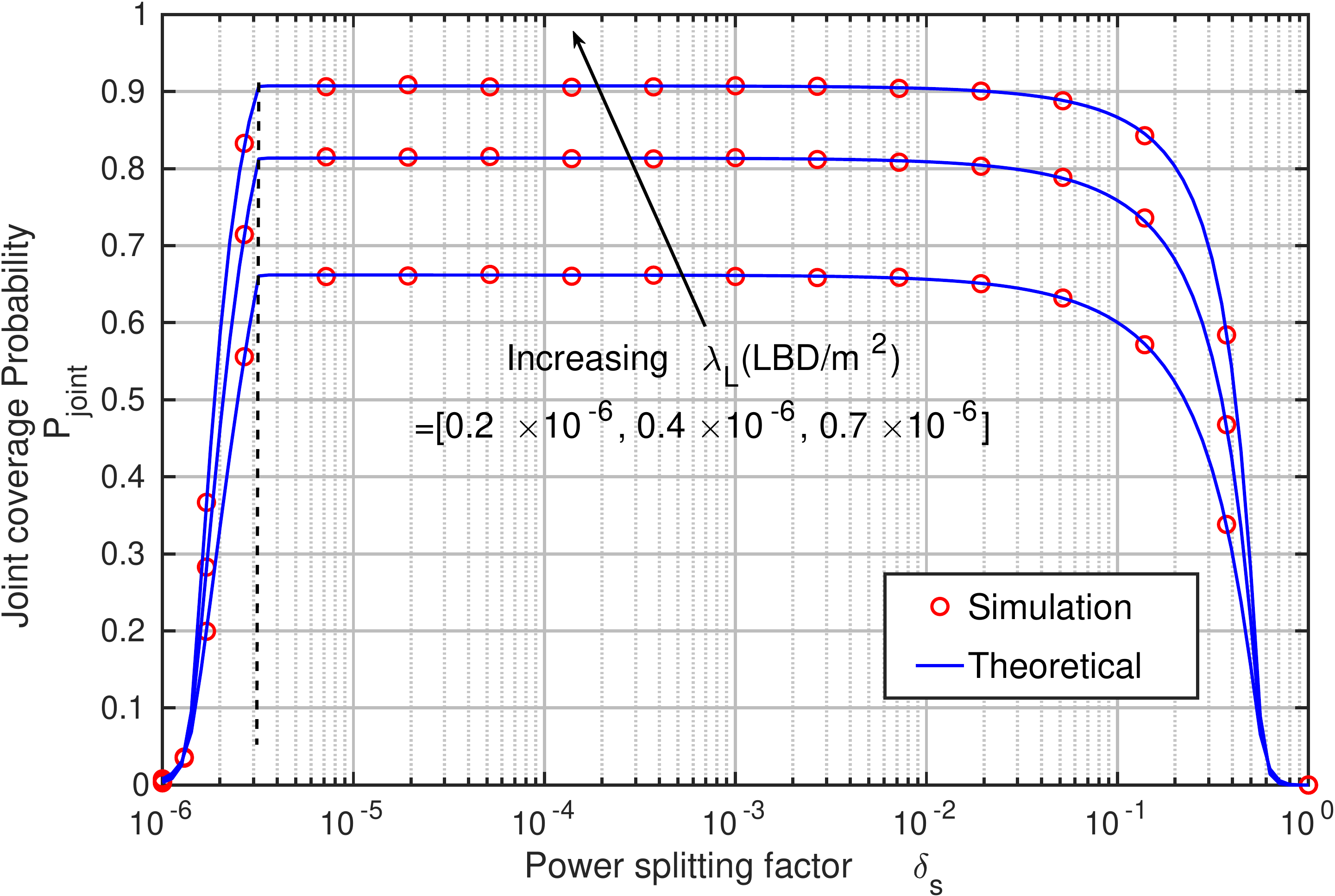}
    \caption{Joint coverage probability against power splitting factor $\delta_s$, $\beta$=55dB. }
    \label{fig:PJOINT2}
\end{figure}

Fig~\ref{fig:PJOINT2} depicts the effect of varying the power splitting factor $\delta_s$ on the joint coverage probability for different values of LBD density $\lambda_L$. We note that at very low values of $\delta_s$, $\mathcal{K}<1$, which makes $P_{\rm joint}$ dominated by SNR coverage and, hence, it is an increasing function of $\delta_s$. However, as $\delta_s$ increases, $P_{\rm joint}$ becomes constant until it starts decreasing as $\delta_s$ approaches $1$.
\section{Conclusion}
In this letter, we used tools from stochastic geometry to study the performance of simultaneous information and power transfer from LBDs to laser-powered UAVs using a power splitting technique. For this setup, we derived the energy coverage probability, the SNR coverage probability, and the joint coverage probability. We demonstrated that, with the current achievable values for the transmission power of the LBD provided by commercially available LBDs, at least 6 LBDs in each $10$ km$^2$, is required to ensure a probability of energy coverage above 0.9. Future work can be oriented towards extending the system by modeling multiple UAVs and studying the optimal scheduling technique when multiple UAVs are within the coverage range of a given LBD.
\section*{Appendix A\\Proof of Lemma~\ref{lem:1}}\label{app:2}
Since $p_{\rm harv}(R)=\frac {(1-\delta_s)\omega A\chi p_{trans} e^{-\alpha \sqrt{H^2+R^2}}}{(D+\sqrt{H^2+R^2}\Delta\theta)^{2}}$ is a strictly decreasing function with respect to the positive radius R, the energy coverage probability is 
\begin{align}
\mathbb{P}( p_{\rm harv}(R)  > p_{\rm prop}+p_{\rm comm})=\mathbb{P}(R  \leqslant R^*)=1-e^{-\lambda_L\pi R^{*2}},
\end{align} 
where $R^*$ is a root for the following equation:
\begin{align}
\label{eq:8}
\frac{(1-\delta_s)\omega A\chi  p_{trans} e^{-\alpha \sqrt{H^2+R^2}}}{(D+\sqrt{H^2+R^2}\Delta\theta)^{2}}= p_{\rm prop}+p_{\rm comm.}\vspace{1.5cm}
 \end{align}

\noindent Let $z=\frac{\alpha}{\Delta\theta}$, and multiply both sides with factor $e^{zD}$, we can rewrite (\ref{eq:8}) as follows:
\begin{small}
 \begin{align}
  (D+\sqrt{H^2+R^2}\Delta\theta)^{2} e^{z(D+\sqrt{H^2+R^2}\Delta\theta)}=\frac{(1-\delta_s)\omega A\chi  p_{trans} e^{zD}}{p_{\rm prop}+p_{\rm comm}}.
 \end{align}
 \end{small}
\noindent As a result, we get an equation with the form of:
\begin{align}\label{slack}
  g(X)=X^2e^{zX}=C,
\end{align}
 \noindent where $z>0$, $C=\frac{(1-\delta_s)\omega A\chi  p_{trans} e^{zD}}{p_{\rm prop}+p_{\rm comm}}>0$ and $X=D+\sqrt{H^2+R^2}\Delta\theta$. \newline To find the solution of (\ref{slack}), let $y(x) = \sqrt{g(x)}=xe^{\frac{zx}{2}} $, $f_1(x)=\frac{zx}{2}$, and $f_2(x)=\frac{2}{z}xe^x$, for every $x>0$. Clearly, we have $y(x)=f_2(f_1(x))$ and $f_1$ is invertible with $f_1^{-1}(x)=\frac{2x}{z}$. However, the inverse of $f_2$ is not straight forward. To find $f_2^{-1}(x)$ we use the Lambert $W$ function defined by the following equation~\cite{corless1996lambertw}:
\begin{align*}
    x=W(x)e^{W(x)}.
\end{align*}
\noindent Consequently the inverse function of $f_2$ is given by:
\begin{align*}
  f_2^{-1}(x)=W_0(\frac{zx}{2}).
\end{align*}
\noindent Recalling that $y(x)=f_2(f_1(x))$, we get
\begin{align*}
    y^{-1}(x)=f_1^{-1}(f_2^{-1}(x))=\frac{2}{z}W_0(\frac{zx}{2}).
\end{align*}
\noindent Having $y^2(x)=g(x)$, we finally get $ g^{-1}(x)=\frac{2}{z}W_0(\frac{z\sqrt{x}}{2})$.
Given that $X^*=g^{-1}(C)=\frac{2}{z}W_0(\frac{z\sqrt{C}}{2})$ and  $X=D+\sqrt{H^2+R^2}\Delta\theta$, the final expression for $R^*$ in (\ref{eqn:r_star}) follows.
\bibliographystyle{IEEEtran}
\bibliography{Draft_v0.11-MK.bbl}

\end{document}